\documentclass[twocolumn,aps,floats,superscriptaddress]{revtex4}
\usepackage{amsmath}

\usepackage{epsfig}
\usepackage{graphicx}
\usepackage{dcolumn}
\usepackage{bm}

\renewcommand{\dag}{^{\dagger}}
\newcommand{\tc}{\tau_{\rm c}}
\newcommand{\Dren}{\Delta_{\rm ren}}
\newcommand{\De}{\Delta}

\def\gapp{\lower.35em\hbox{$\stackrel{\textstyle>}{\sim}$}}
\def\lapp{\lower.35em\hbox{$\stackrel{\textstyle<}{\sim}$}}

\begin{document}
%

\title{
Entanglement and dephasing of quantum dissipative systems}
\author{T. Stauber} 
\author{F. Guinea}
\affiliation{Instituto de Ciencia de Materiales de Madrid, CSIC, Cantoblanco, E-28049 Madrid, Spain.}
\date{\today}
\begin{abstract}
The von Neumann entropy of various quantum dissipative models is calculated
in order to discuss the entanglement properties of these systems. First,
integrable quantum dissipative models are discussed, i.e., the quantum
Brownian motion and the quantum harmonic oscillator. In case of the free
particle, the related entanglement of formation shows no non-analyticity. In
case of the dissipative harmonic oscillator, 
there is a non-analyticity at the transition of underdamped to overdamped
oscillations. We argue that this might be a general property of dissipative
systems. We show that similar features arise in the dissipative two level
system and study different regimes using sub-Ohmic, Ohmic and
and super-Ohmic baths, within a scaling approach.
\end{abstract}
%

%
%
%
%
\maketitle 
\section{Introduction}
The operation of a quantum computer requires a careful control of the
interaction between the system and its macroscopic environment. The resulting
entanglement between the system's degrees of freedom and the reservoir has
been a recurrent topic since the formulation of quantum mechanics, as it
is relevant to the analysis of the measurement process \cite{O92,Z03}. The 
loss of coherence due to the interaction of a quantum system and its
environment was also extensively studied due to its importance to Macroscopic
Quantum Tunneling and related effects \cite{CL81}. Theoretical research on
Macroscopic Quantum Tunneling lead, among other results, to the formulation
of a canonical model for the analysis of a quantum system interacting with a
macroscopic environment, the so called Caldeira-Leggett model \cite{Cal83}. It
can be shown that this harmonic model (see below) describes correctly the low
energy features of a system which, in the classical limit, undergoes ohmic
dissipation (linear friction). It can be extended to systems with more
complicated, non linear, dissipative properties (usually called sub-Ohmic and
super-Ohmic, see below) \cite{Letal87,W99}.

In relation to the ongoing research on entanglement, a recent interesting
development is the analysis of the enhancement of entanglement in a system
near a quantum critical point \cite{Oetal02,ON02}.
The original systems under study were the transverse Ising model and the XY
model, but also other models which exhibit a quantum phase transition were 
later investigated in this direction, as e.g. the Lipkin-Meshkov-Glick
model \cite{Vidal04,DV05}.  

A connection between previous research on Macroscopic Quantum Tunneling and
entanglement near quantum critical points is starting to
emerge \cite{SG04}. It is interesting to remark that some of the simplest
systems which show a non trivial quantum critical point is the dissipative
two level system \cite{Letal87} and related models, like the Kondo
model \cite{AYH70}. Similar models describe quantum fluctuations in Josephson
junctions \cite{S83} or tunneling between Luttinger liquids \cite{KF92}. It is
already known that, even for the ground state of simple models like the dissipative harmonic oscillator non-trivial entanglement properties can be expected as
was already commented on in Ref. \cite{NB02}. Entanglement energetics at zero 
temperature was investigated in Ref. \cite{JB04}. A number of properties of
the entanglement in the Caldeira-Leggett model and related models remain,
however, unexplored.

The models studied here describe a quantum system characterized by a small
number of degrees of freedom coupled to a macroscopic reservoir. 
These models show
a crossover between different regimes, or even exhibit a quantum
critical point. As this behavior is induced by the presence
of a reservoir with a large number of degrees of freedom, they can also be
considered as a model of dephasing and loss of quantum coherence. It is worth
noting that there is a close connection between models describing 
impurities coupled to a reservoir, and strongly correlated systems 
near a quantum critical point, as evidenced by Dynamical Mean Field 
Theory \cite{GKKR96}. In the limit of large coordination, the properties 
of an homogeneous system can be reduced to
those of an impurity interacting with an appropriately chosen
reservoir. Hence, in the limit of large coordination
the entanglement between the quantum system and the reservoir near a phase
transition can be mapped onto the entanglement which develops in an
homogeneous system  near a quantum critical point.

The measure of entanglement used in the original papers is the the
concurrence  introduced by Wooters \cite{W98} Alternatively, the von Neuman 
entropy of macroscopic (contiguous) subsystems can be used \cite{VPC04}. 
A non-local measure of entanglement was employed in the study of the 
Affleck-Kennedy-Lieb-Tasaki (AKLT) model \cite{Vetal03,VMC04}.  

Our previous work \cite{SG04} showed that the main
non-analyticity of the concurrence arises at the transition of coherent to
incoherent tunneling. At the actual quantum phase transition, we only found a
much weaker non-analyticity associated with the existence of the Kosterlitz-Thouless
weakly non analytical features. In the same way, the phase transition of the transverse
Ising model discussed in Refs. \cite{Oetal02,ON02} can be interpreted as a
transition where coherence is lost due to the emergence of a localized state
at the transition. 
We thus assume that the loss of coherence might be more important to see
non-analyticities in the entanglement of a system than the actual phase transition.

In the first part of this article, we test our assumption using two
integrable quantum dissipative models, 
the dissipative free particle - that is, the Caldeira-Leggett model - and the
dissipative quantum harmonic oscillator. 
These models do not exhibit a quantum phase transition, but, in the latter
case there is a transition 
from underdamped to overdamped oscillations at some critical coupling
strength. 
As measure of entanglement we use the von Neuman entropy of the subsystem,
defined using the reduced density matrix, $\rho_A$, obtained by tracing out
the bath degrees of freedom 
of the ground state:
\begin{align}
E(\psi)=-\text{Tr}(\rho_A\ln\rho_A)\quad,\quad\rho_A=\text{Tr}_B(|\psi\rangle\langle\psi|)
\label{entropy}
\end{align}

In the second part of the paper, we make the same analysis for the spin-boson model on the
basis of a scaling approach for the free energy. 
For super-Ohmic baths, the model shows no phase transition whereas for Ohmic
and sub-Ohmic baths, 
there is a transition from localized to non-localized behavior. 
Again, we focus the discussion on the transition from coherent to 
incoherent oscillation which exists for Ohmic dissipation, but is also present for certain non-Ohmic environments. 
 
It is finally worth noting that a more mathematical analysis of some problems
related to the entanglement in the dissipative harmonic oscillator can be found in Refs. \cite{EP02,AEPW02}.
Sub-Ohmic environments may be relevant to the description no Gaussian effects
in qubits coupled to external environments, see \cite{PFFF02,FDMP04}.

\section{Exactly solvable dissipative systems}
Modeling the environment by a set of harmonic oscillators \cite{Cal83}, the general integrable model is described by the following Hamiltonian:
\begin{align}
\label{HOscillator}
    H&=\frac{p^2}{2}+\frac{\omega_0^2}{2} q^2+
     \sum_{\alpha}\Big(\frac{p_{\alpha}^2}{2}
        +\frac{1}{2}
        \omega_{\alpha}^2\big(x_{\alpha}-\frac{\lambda_{\alpha}}{\omega_{\alpha}^2}q\big)^2\Big)
\end{align}
The operators obey the canonical commutation relations which
read ($\hbar=1$)
\begin{align}
 \left[q,p\right]=i\quad,\quad\left[x_{\alpha},p_{\alpha^{\prime}}\right]=i
 \delta_{\alpha,\alpha'}\quad.
\end{align}
The coupling of the system to the bath is completely determined by the spectral function
\begin{align}
J(\omega)=\frac{\pi}{2}\sum_\alpha\frac{\lambda_\alpha^2}{\omega_\alpha}.
\end{align}
In the following, we will consider a Ohmic bath with $J(\omega)=\eta\omega$ for $\omega\ll\omega_c$ and $J(\omega)=0$ for $\omega\gg\omega_c$, $\omega_c$ being the cutoff frequency.
\subsection{Caldeira-Leggett model}
Let us first consider the free dissipative particle, i.e., we set $\omega_0=0$. The model was introduced by Caldeira and Leggett and further investigated by Hakim and Ambegaokar \cite{Cal83b,Hak85}. The latter authors obtained the reduced density matrix via diagonalization of the Hamiltonian:
\begin{align}
\langle x|\rho_A|x'\rangle=e^{-a(x-x')^2}/L\quad,\quad a=\frac{1}{4}\frac{\eta}{\pi}\ln\left(1+\frac{\omega_c^2}{\eta^2}\right)
\label{densitymatrix}
\end{align}
where $\eta$ denotes the friction coefficient and $\omega_c$ is the cutoff frequency of the bath. Furthermore, $L\to\infty$ denotes the system size and in contrary to the use of Eq. \ref{densitymatrix} in Ref. \cite{Hak85}, here the normalization is crucial to assure Tr$\rho_A=1$. 

In order to calculate the entropy of the system, we Taylor expand the logarithm:
\begin{align}
\label{LogEx}
\ln\rho_A=-\sum_{n=1}\frac{(1-\rho_A)^n}{n}=-\sum_{n=1}\frac{1}{n}\sum_{k=0}^n{n\choose k}(-1)^k\rho_A^k
\end{align} 
Further we have
\begin{align}
\langle x|\rho_A^k|x'\rangle=\sqrt{\frac{\pi}{a}}^{k-1}\sqrt{\frac{1}{k}}e^{-\frac{a}{k}(x-x')^2}/L^k
\end{align}
proved by induction. With the identity
\begin{align}
\sqrt{\frac{1}{k}}=\frac{1}{\sqrt{\pi}}\int dxe^{-kx^2}
\end{align}
we thus obtain for the specific entropy (for general dimension $d$)
\begin{align}
S=\frac{d}{2}\left(\ln(aL^2)+1-\ln\pi\right).
\end{align}
Comparing the above result with the entropy of a particle in a canonical ensemble, we identify $a\sim\lambda^{-2}\propto T$ with $\lambda$ denoting the thermal de Broglie wavelength and $T$ the temperature of the canonical ensemble.

Notice that the entropy of a free dissipative particle shows no non-analyticity.
\subsection{Dissipative harmonic oscillator}
We now include the harmonic potential, i.e., $\omega_0\neq0$.
The reduced density matrix of the damped harmonic oscillator is given by \cite{W99}
\begin{align}
\langle x|\rho_A&|x'\rangle=\sqrt{\frac{4b}{\pi}}e^{-a(x-x')^2-b(x+x')^2}\\
&a=\frac{\langle p^2\rangle}{2}\quad,\quad b=\frac{1}{8\langle q^2\rangle}\;.\notag
\end{align}
The above expression is deduced such that the correct variances for position and momentum are obtained. At $T=0$ the expectation values are given by
\begin{align}
\langle q^2\rangle&=\frac{1}{2\omega_0}f(\kappa)\\
\langle p^2\rangle&=\omega_0^2(1-2\kappa^2)\langle q^2\rangle+\frac{2\omega_0\kappa}{\pi}\ln\left(\frac{\omega_c}{\omega_0}\right)
\end{align} 
with $\kappa=\eta/2\omega_0$ and 
\begin{align}
f(\kappa)=\frac{1}{\pi}\frac{\ln\left[(\kappa+\sqrt{\kappa^2-1})/(\kappa-\sqrt{\kappa^2-1})\right]}{\sqrt{\kappa^2-1}}\quad.
\end{align}
The parameter $\kappa$ represents the friction parameter and the system experiences a crossover from coherent to incoherent oscillations for $\kappa=1$.

Taylor expanding the logarithm of the entropy, Eq. (\ref{LogEx}), we need to evaluate the general $n$-dimensional integral
\begin{align}
\int_{-\infty}^{\infty}dx_1..dx_n\exp\left(-\sum_{i,j=1}^nx_iA_{i,j}x_j\right)=\frac{\pi^{n/2}}{\sqrt{\text{det}A}}
\end{align}
where $A$ is given by the translationally invariant tight-binding matrix with $A_{i,i}=2(a+b)$, $A_{i+1,i}=A_{i,i+1}=-(a-b)$ ($n+1\equiv1$) and zero otherwise. The determinant of the matrix is given by its eigenvalues and reads
\begin{align}
\text{det}A=(2a)^n(1-b/a)^n\prod_{m=1}^n\left[1+\frac{2b}{a-b}-\cos k_m\right]
\end{align}
with $k_m=2\pi m/n$. Considering the $n$-dimensional translationally invariant, but non-hermitian matrix $\widetilde A_{i,i}=1$, $\widetilde A_{i+1,i}=1-\varepsilon$ ($n+1\equiv1$) and zero otherwise, one obtains the following formula:
\begin{align}
\prod_{m=1}^n\left[1+\frac{\varepsilon^2}{2(1-\varepsilon)}-\cos k_m\right]=\frac{(1-(1-\varepsilon)^{n})^2}{2^n(1-\varepsilon)^n}
\end{align}
For $\omega_c/\omega_0\gg1$, we have
\begin{align}
a/b&=4\langle q^2\rangle\langle p^2\rangle\notag\\
&=f(\kappa)\left[(1-2\kappa^2)f(\kappa)+\frac{4\kappa}{\pi}\ln\left(\frac{\omega_c}{\omega_0}\right)\right]\gg1.
\end{align}
In this limit, we can thus set $\varepsilon^2=4b/a\ll1$ and the $n$-dimensional integral can be approximated to yield
\begin{align}
\int dx\langle x|\rho_A^n|x\rangle&\rightarrow\frac{\tilde\varepsilon^n}{1-(1-\varepsilon)^n}\;,
\end{align}
with $\tilde\varepsilon\equiv\varepsilon\sqrt{1-\varepsilon}/\sqrt{1-\varepsilon^2/4}$.
Expanding the denominator as geometrical series,
we have for the entropy
\begin{align}
S&=-\left(\frac{\tilde\varepsilon}{\varepsilon}\ln\tilde\varepsilon+\frac{\tilde\varepsilon}{\varepsilon^2}\ln(1-\varepsilon)\right).
\label{entropieO}
\end{align}

\begin{figure}[t]
  \begin{center}
    \includegraphics*[width=3.5in,angle=0]{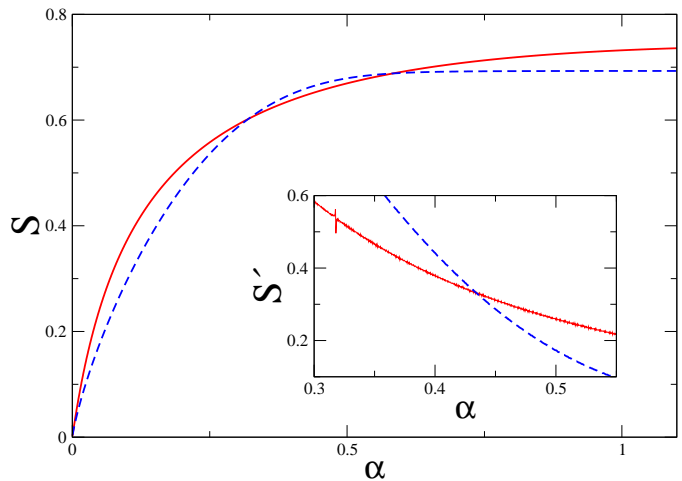}
    \caption{(Color online). The entropy $S$ of the dissipative oscillator (full) and the dissipative two-level system (dashed) with Ohmic coupling as function of the dimensionless coupling strength $\alpha$. Inset: The derivative of the entropy $S'$ with respect to the coupling strength $\alpha$ is shown as function of $\alpha$.}
    \label{GEntropie}
\end{center}
\end{figure}

In the limit $\varepsilon\approx\tilde\varepsilon\ll1$, the leading behavior of the entropy is given by $S\sim\ln(a/b)$. We thus find a non-analyticity at $\kappa=1$, the point of the crossover of incoherent to coherent oscillations. The leading behavior of Eq. (\ref{entropieO}) is plotted in Fig. \ref{GEntropie} as function of the dimensionless coupling strength $\alpha=q_0^2\eta/(2\pi)$ with the characteristic length scale $q_0=1/\sqrt{\omega_0}$ for $\omega_c/\omega_0=100$ (full line). In the inset, the non-analyticity of the derivative of the entropy with respect to the coupling strength $S'\equiv \partial_\alpha S$ at $\alpha=1/\pi$ can be seen. 
\section{Spin-Boson Model}
A prominent dissipative model is given by the spin-boson model or dissipative two-level system (TLS). The Hamiltonian without bias reads
\begin{align}
    H=\frac{\De_0}{2}\sigma_x  
        +\sum_k\omega_{k}b_{k}\dag b_k 
        +\sigma_z\sum_k\frac{\lambda_{k}}{2}(b_k+b_k\dag)\quad.
\end{align}
The operators $b_{k}^{(\dagger)}$ resemble the bath degrees of freedom
and $\sigma_x$, $\sigma_y$, $\sigma_z$ denote the Pauli spin matrices. They
obey the canonical commutation relations and the spin-$1/2$
algebra, respectively.

The coupling constants $\lambda_k$ are parameterized by the spectral function
\begin{align}
J(\omega)=\sum_k\lambda_k^2\delta(\omega-\omega_k).
\label{spectralFunctionSB}
\end{align}
In the relevant low-energy regime, the spectral function is generally parameterized as a power-law, i.e., $J(\omega)\propto2\alpha\omega^s\Lambda_0^{1-s}$ where $\alpha$ denotes the coupling constant, $s$ the bath type and  $\Lambda_0$ the cutoff-frequency. 

The general reduced density matrix of the spin-boson model is given by
\begin{align}
\rho_A=\frac{1}{2}
\begin{pmatrix}
1 + \langle
\sigma_z \rangle &\langle \sigma_x \rangle \\ \langle \sigma_x
\rangle &1 - \langle \sigma_z \rangle
\end{pmatrix}.
\label{entropySB}
\end{align}
Since there is no symmetry breaking field in the above Hamiltonian, we have $\langle \sigma_z\rangle=0$. The eigenvalues are thus given by $\lambda_\pm=(1\pm\langle \sigma_x \rangle)/2$ and the entropy reads
\begin{align}
S=-\frac{1}{2}\left[\ln\left((1-\langle \sigma_x \rangle^2\right)/4)+\langle \sigma_x \rangle\ln\left(\frac{1+\langle \sigma_x \rangle}{1-\langle \sigma_x \rangle}\right)\right].
\end{align}
The value of $\langle \sigma_x \rangle$, at zero
temperature, is given by 
\begin{equation}
\langle \sigma_x \rangle = 2\frac{\partial E}{\partial \Delta_0}
\label{der_ener}
\end{equation}
where $E$ is the energy of the ground-state. To obtain the ground-state energy, a scaling analysis for the free energy at arbitrary temperature is considered, see the appendix. In the following, we use this approach to calculate $E ( \Delta_0 )$ and $\langle \sigma_x \rangle$ which will set the basis of our discussion on the entanglement properties of the spin-boson model.

\subsection{Ohmic dissipation}
In the Ohmic case ($s=1$), there is a phase transition at zero temperature at the critical coupling strength $\alpha=1$ \cite{Bra82,Cha82}. The transition is also manifested in the renormalized tunnel element $\Dren$, i.e., $\Dren=\De_0(\De_0/\Lambda_0)^{\alpha/(1-\alpha)}$ for $\alpha<1$ and $\Dren=0$ for $\alpha>1$. 

The free energy is determined by (see the appendix)
\begin{equation}
F = \int_{\Dren}^{\Lambda_0} \left( \frac{\Delta ( \Lambda )}{\Lambda}
\right)^2 d \Lambda\;.
\label{int_ener}
\end{equation}
The ground state energy $E$ can then be written as
\begin{equation}
E = \left\{ \begin{array}{lr} \frac{C}{1 - 2
\alpha} \left[ \Delta_0 \left( \frac{\Delta_0}{\Lambda_0}
\right)^{\frac{\alpha}{1 - \alpha}} - \frac{\Delta_0^2}{\Lambda_0}
\right] &0 < \alpha < \frac{1}{2} \\ 2 C
\frac{\Delta_0^2}{\Lambda_0} \log \left( \frac{\Lambda_0}{\Delta_0}
\right) &\alpha = \frac{1}{2}
\\ \frac{C}{2 \alpha - 1} \left[ \frac{\Delta_0^2}{\Lambda_0} - \Delta_0
\left( \frac{\Delta_0}{\Lambda_0} \right)^{\frac{\alpha}{1 - \alpha}}
\right] &\frac{1}{2} < \alpha < 1 \\ C
\frac{\Delta_0^2}{\Lambda_0} & \alpha > 1
\end{array} \right. \label{sx1/2_TLS}
\end{equation}
where $C$ is a numerical constant. For $\alpha=1/2\pm\epsilon$, we have
\begin{equation}
\frac{d\ln\langle\sigma_x\rangle}{d\alpha}\Big|_{\alpha=1/2\pm\epsilon}\propto\frac{1}{\epsilon}.
\label{Divergence}
\end{equation}
For $\alpha=1-\epsilon$, we have
\begin{equation}
\frac{d\ln\langle\sigma_x\rangle}{d\alpha}\Big|_{\alpha=1-\epsilon}\propto \ln\left(\frac{\Delta_0}{\Lambda_0}\right)
\end{equation}
The non-analyticity around $\alpha=1$ is thus far weaker than around $\alpha=1/2$. This non-analyticity is also present in the entropy as can be seen from the expression Eq. (\ref{entropySB}).

The entropy $S$ of the dissipative two-level system with Ohmic coupling is plotted in Fig. \ref{GEntropie} as function of the dimensionless coupling strength $\alpha$ for $\omega_c/\Delta_0=100$ (dashed line). The inset shows the derivative of the entropy with respect to the coupling strength. The entropy quickly saturates after the transition from coherent to incoherent oscillations at $\alpha=1/2$, but the non-analyticity of Eq. (\ref{Divergence}) cannot be seen on this scale.
\subsection{Non-Ohmic dissipation})
The calculation of $E ( \Delta_0 )$ and $\langle \sigma_x \rangle$ can be
extended to the spin-boson model with non-Ohmic dissipation ($s\neq1$). In general, the dependence of the effective tunneling term on the cutoff, $\Delta( \Lambda )$ is:
\begin{equation}
\Delta ( \Lambda ) = \Delta_0 \exp\left(- \frac{1}{2}\int_\Lambda^{\Lambda_0} \frac{J (
    \omega )}{\omega^2} d \omega\right) \label{delta_l}
\end{equation}
with the spectral function given in Eq. (\ref{spectralFunctionSB}). A renormalized low energy term, $\Dren$, can be defined by
\begin{equation}
\Dren = \Delta_0 e^{- \int_{\Dren}^{\Lambda_0} \frac{J (
    \omega )}{\omega^2} d \omega}\;.
\label{Dren}
\end{equation}
The free energy is again determined by Eq. (\ref{int_ener}), though cannot be
evaluated analytically, anymore. The scaling behavior of the renormalized tunneling given in Eq.(\ref{delta_l}) is no longer a power
law, as in the Ohmic case. Still, we can distinguish two limits:

i) The renormalization of $\Delta ( \Lambda )$ is slow. In this case,  the integral
in Eq. (\ref{int_ener}) is dominated by the region $\Lambda \sim \Lambda_0$,
where the function in the integrand goes as $\Lambda^{-2}$. The integral is
dominated by its high cutoff, $\Lambda_0$, and the contribution from the
region near the lower cutoff, $\Dren$, can be neglected. Then, we obtain that
 $F ( \Delta_0 ) \sim \Delta_0^2 / \Lambda_0$.

ii) The renormalization of  $\Delta ( \Lambda )$ is fast. In this case, the contribution to the integral 
in Eq. (\ref{int_ener}) from the region $\Lambda \approx \Lambda_0$ is
small. The value of the integral is dominated by the region near $\Lambda
\simeq \Dren$. As $\Dren$ is the only quantity with dimensions of energy
needed to describe the properties of the system in this range, we expect that $F
( \Delta_0 ) \approx \Dren$.

In the scaling limit, $\Delta_0 /
\Lambda_0 \ll 1$, the values of the two terms, $\Dren$
and $\Delta_0^2 / \Lambda_0$, become very different. In
addition, there are no other energy scales which can qualitatively modify the
properties of the system. We thus conclude that only the two terms mentioned above will contribute to the free energy. Hence, we can write:
\begin{equation}
F ( \Delta_0 ) \simeq {\rm max} \left( \Dren , \frac{\Delta_0^2}{\Lambda_0}
\right)
\label{freeEnergy}
\end{equation}

The above equation is now used to discuss the possible transition between underdamped to overdamped oscillations for non-Ohmic environments. Notice that it also applies for Ohmic baths.
\subsubsection{Super-Ohmic dissipation}
In the super-Ohmic case ($s > 1$), Eq. (\ref{Dren}) always has a solution and,
moreover, we can also set the lower limit of the integral to zero. This yields
\begin{equation}
\Dren = \Delta_0 e^{- \int_0^{\Lambda_0} \frac{J (
    \omega )}{\omega^2} d \omega} \approx \Delta_0 e^{- \alpha / ( s- 1 )}. 
\end{equation}
For $\alpha \gg 1$ we have $\Dren \ll \Delta_0$, but there is no transition from localized to delocalized behavior. 

Using Eq. (\ref{freeEnergy}) in the super-Ohmic case $s > 1$, we can approximately
write:
\begin{equation}
\langle \sigma_x \rangle \simeq {\rm max} \left( e^{- \alpha / ( s-1)} ,
  \frac{\Delta_0}{\Lambda_0} \right)
\end{equation}
We thus find a transition from underdamped to overdamped oscillations at some
critical coupling strength $\alpha \sim (s-1) \log ( \Lambda_0 / \Delta_0 )$.

It is finally interesting to note that the scaling analysis discussed
in Ref. \cite{K76} is equivalent to the scheme used here.
\subsubsection{Sub-Ohmic dissipation}
In the sub-Ohmic case ($s < 1$), it is not guaranteed that Eq. (\ref{Dren}) has a solution. In general, a solution only exists when $\Delta_0 / \Lambda_0$ is
not much smaller than 1. 

The existence of a phase transition in case of a sub-Ohmic bath was first proved in Ref. \cite{Spohn85}. Whereas the relation in Eq. (\ref{Dren}) and a similar analysis based on flow equations for Hamiltonians \cite{KM96} yields a discontinuous transition between the localized and delocalized
regimes, detailed numerical calculations suggest that the transition is continuous \cite{Bulla03}. 

Since there is a phase transition from localized to non-localized behavior,
there might also be a 
transition between overdamped to underdamped oscillation. 
In Ref. \cite{SM02}, this transition was discussed on the basis of spectral 
functions analogous to the discussion of Ref. \cite{G85,Costi96} for Ohmic
dissipation. 
It was found that for $s>0.5$ the transition takes place for lower 
values of $\alpha$ as in the Ohmic case, e.g., for $s=0.8$ and $\Lambda_0/\Delta_0=10$ the transition coupling strength is $\alpha^*\approx0.2$. 

Using Eq. (\ref{Dren}) and Eq. (\ref{freeEnergy}) yields for the sub-Ohmic case:
\begin{equation}
\langle \sigma_x \rangle \simeq \left\{ \begin{array}{llr} 1 &{\rm delocalized
      \, \, \,
      regime \, \, \,} &\frac{\Delta_0}{\Lambda_0} \simeq 1  \\
      \frac{\Delta_0}{\Lambda_0} &{\rm localized \, \, \, regime \, \, \,} 
      &\frac{\Delta_0}{\Lambda_0} \ll 1
      \end{array} \right.
\end{equation}
The analysis used in the previous cases leads us to expect coherent
oscillations in the delocalized regime.

We can extend the study of the sub-Ohmic case to the vicinity of the second
order transition described in Ref. \cite{VTB05}, which in our notation takes place for $\alpha=s\Delta_0 / \Lambda_0\ll 1$. In this regime, which
cannot be studied using  the Franck-Condon like renormalization in
Eq. (\ref{Dren}), we use the renormalization scheme around the fully coherent
state proposed in Ref. \cite{VTB05}. To one-loop order, the beta-function for the dimensionless
quantity (expressed in our notation) $\tilde{\kappa} = ( \alpha \Lambda ) / \Delta$ then reads 
\begin{equation}
\beta(\tilde{\kappa}) = -s\tilde{\kappa}+\tilde{\kappa}^2.
\label{beta_k}
\end{equation}
Near the transition, in the delocalized phase, $\tilde{\kappa}$ thus scales towards zero as
\begin{equation}
\tilde{\kappa} ( \Lambda ) = \tilde{\kappa}_0 \left(
  \frac{\Lambda}{\Lambda_0} \right)^s\;.
\label{scaling_k}
\end{equation}
The scaling of $\langle \sigma_x \rangle$ is
\begin{equation}
\frac{\partial \langle \sigma_x \rangle}{\partial \Lambda} = - \tilde{\kappa}
( \Lambda ) \frac{\Delta}{\Lambda^2}\;.
\label{scaling_sx}
\end{equation}
The fact that the scheme assumes a fully coherent state as a starting point
implies that $\Delta$ is not renormalized. Inserting Eq. (\ref{scaling_k})
into Eq. (\ref{scaling_sx}), we find:
\begin{equation}
\frac{\partial \langle \sigma_x \rangle}{\partial \Lambda} = -  \tilde{\kappa}_0 \left(
  \frac{\Lambda}{\Lambda_0} \right)^s \frac{\Delta_0}{\Lambda^2}
\end{equation}
\begin{figure}[t]
  \begin{center}
    \includegraphics*[width=3.5in,angle=0]{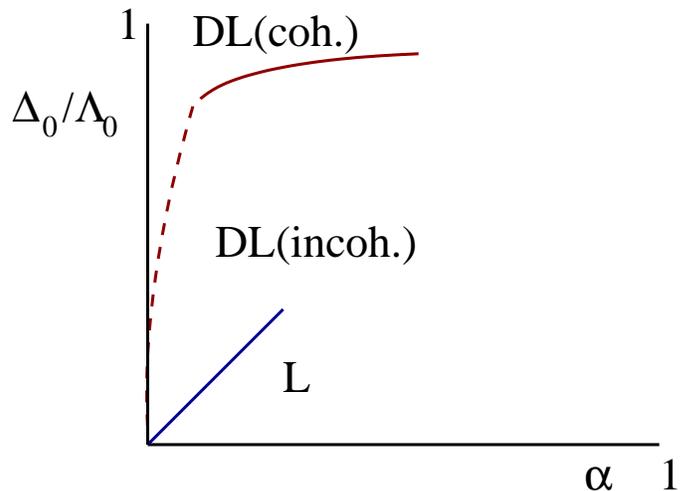}
    \caption{(Color online). Schematic picture of the different regimes in the sub-Ohmic
      dissipative TLS studied in the text. DL stands for delocalized phase,
      while L denotes a localized phase. The lower blue line denotes the
      continuous transition studied in Ref. \protect{\cite{VTB05}}. The red line
      marks the boundaries of a regime characterized by a small
      renormalization of the tunneling rate, Eq. (\protect{\ref{Dren}}), and
      coherent oscillations.}
    \label{subohmic}
\end{center}
\end{figure}

If we calculate $\langle \sigma_x \rangle$ from this equation, we find
that the resulting integral diverges as $\Lambda \rightarrow 0$ for $s \leq
1$. This result implies that $\langle \sigma_x \rangle \ll 1$. For
sufficiently low values of the effective cutoff, $\Lambda$, the value of
$\langle \sigma_x \rangle$ can be calculated using a perturbation expansion
on $\Delta_0$, leading to $\langle \sigma_x \rangle \sim \Delta_0 /
\Lambda_0$. This result implies the absence of coherent oscillations,
as in the similar cases discussed previously.

A schematic picture of the regimes studied for the sub-Ohmic TLS is shown in
Fig. [\ref{subohmic}]. 
\section{Summary}
In this article, the entanglement properties of dissipative systems were investigated on the basis of the von Neumann entropy.

We first investigated two integrable dissipative quantum systems -the free dissipative particle and the dissipative harmonic oscillator - and calculated the von Neumann entropy. In the former case, this could be done exactly and no non-analyticity was found. The case of the harmonic oscillator is the more interesting one since it exhibits a transition from underdamped to overdamped oscillations for increasing dissipation. This transition is also manifested in the entropy, or equivalently in the entanglement which was calculated in the limit of large bath cutoff.

We also calculated the von Neumann entropy for the spin-boson model on the basis of a scaling equation for the free energy. Only in the Ohmic case, the resulting integral could be evaluated and we analyzed the non-analyticity at the transition from underdamped to overdamped oscillations. We found that the non-analyticity more pronounced than at the actual phase transition.     

In the non-Ohmic case, we argued that the transition between coherent and
decoherent oscillation takes place when 
the value of $\langle \sigma_x \rangle$ becomes comparable to the result
obtained using a perturbation expansion in the tunneling matrix, $\Delta$ (as
is the case for Ohmic dissipation). 
In the super-Ohmic case, this always yields a critical coupling strength at
zero temperature which differs from the analysis in Ref. \cite{Leggett87}. 

In the sub-Ohmic case, the scaling approach can only be trusted when the tunnel matrix element is of the same order of magnitude as the cutoff. Then a transition between coherent to non-coherent oscillations is possible before the system becomes localized. For the regime where the cutoff represents the largest energy scale, we applied a novel renormalization scheme proposed in Ref. \cite{VTB05}. We find that, in the delocalized phase, the system is most likely incoherent.  

Concerning the entanglement properties for the non-Ohmic case, we were not able to discuss possible non-analyticities since the regime is analytically not accessible. Numerical work in this direction is planned for the future.

To conclude, we suppose that entanglement properties are closely connected to the transition of coherent to incoherent tunneling. Our observations might be useful for future quantum bit manipulations.  
\section{Acknowledgments}
Funding from MCyT (Spain) through ``Juan-de-la-Cierva'' and grant MAT2002-04095-C02-01 is acknowledged.
\appendix
\section{Calculation of the free energy of the dissipative TLS}
We calculate the free energy of the dissipative two level system following the
scaling approach discussed for the Kondo problem in Refs. \cite{AYH70,AY71}, and formulated in a more general way in Ref. \cite{C81}. For the general long-ranged Ising model, the scaling approach was first applied by Kosterlitz \cite{K76}.

\begin{figure}[t]
  \begin{center}
    \includegraphics*[width=3in,angle=0]{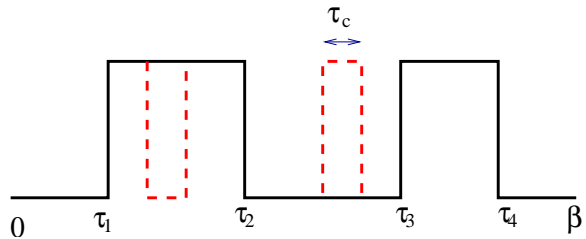}
    \caption{(Color online). Sketch of the instanton pairs which renormalizes the calculation
      of the free energy of the dissipative TLS.}
    \label{free_energy}
\end{center}
\end{figure}

The partition function of the model can be expanded in powers of $\Delta^2$ as
\begin{equation}
Z = \sum_n \frac{\Delta^{2n}}{2n !} \int_0^\beta d \tau_1 
\cdots \int_0^\beta d
\tau_{2n} \prod_{ij=1,..,2n} f [ ( \tau_i - \tau_j ) / \tc ]
\label{free_ener}
\end{equation}
where $f [ ( \tau_i - \tau_j ) / \tc ]$ denotes the interaction between the kinks
located at positions $\tau_i$ and $\tau_j$. A term in the series is
schematically depicted in Fig. [\ref{free_energy}]. The scaling procedure
lowers the short time cutoff of the theory from $\tc$ to $\tc - d \tc$. This process
removes from each term in the sum in Eq. (\ref{free_ener}) details at times
shorter than $\tc - d \tc$. The rescaling $\tc \rightarrow \tc - d \tc$
implies the change $\Delta \rightarrow \Delta ( 1 + d \tc / \tc )$. The
dependence of $f [ ( \tau_i - \tau_j ) / \tc ]$ leads to another rescaling,
which can be included in a global renormalization of
$\Delta$ \cite{AYH70,AY71,C81}. In addition, configurations with an
instanton-antiinstanton pair at distances between $\tc$ and $\tc - d \tc$
have to be replaced by configurations where this pair is absent, as
schematically shown in Fig. [\ref{free_energy}]. The number of removed pairs
is proportional to $d \tc / \tc$. The center of the pair can be anywhere in
the interval $0 \le \tau \le \beta$. The final effect is the rescaling:
\begin{equation}
Z \rightarrow Z \left( 1 + \Delta^2 \beta d \tc \right)
\label{ren_Z}
\end{equation}
Writing $Z$ as $Z = e^{- \beta F}$, where $F$ is the free energy,
Eq. (\ref{ren_Z}) can be written as:
\begin{equation}
- \frac{\partial F}{\partial \tc} = \Delta^2 ( \tc )
\end{equation}
In the Ohmic case, the dependence of $\Delta$ on $\tc = \Lambda^{-1}$ is
\begin{equation}
\Delta ( \Lambda ) = \Delta_0 \left( \frac{\Lambda}{\Lambda_0} \right)^\alpha
\end{equation}
and, finally, we find the following relation:
\begin{equation}
\frac{\partial F}{\partial \Lambda} =  \left[ \frac{\Delta ( \Lambda
    )}{\Lambda} \right]^2 = \left( \frac{\Delta_0}{\Lambda_0} \right)^2
    \left( \frac{\Lambda}{\Lambda_0} \right)^{2 \alpha - 2}
\end{equation}
This equation ceases to be valid for $\Lambda \simeq \Delta_{\rm
  ren}$. For finite temperatures, we obtain
\begin{equation}
F ( T ) = \int_T^{\Lambda_0} \frac{\partial F}{\partial \Lambda} d \Lambda\;.
\label{int_free}
\end{equation}

It is interesting to apply this analysis to a free two level system. The
value of $\Delta_0$ does not change under scaling. We find the following expression:
\begin{equation}
\frac{\partial F}{\partial \Lambda} = \left\{ \begin{array}{lr} \left( \frac{\Delta_0}{\Lambda}
\right)^2 &\Delta_0 \ll \Lambda \\ 0 &\Lambda \ll \Delta_0 \end{array} \right.
\end{equation}
Inserting this expression into Eq. (\ref{int_free}), we obtain
\begin{equation}
F ( T ) = \left\{ \begin{array}{lr} \frac{\Delta_0^2}{T} &\Delta_0 \ll T \\
    \Delta_0 &T \ll \Delta_0 \end{array} \right.
\end{equation}
and, finally:
\begin{equation}
\langle \sigma_x \rangle = \frac{\partial F}{\partial \Delta_0} = 
\left\{ \begin{array}{lr} \frac{\Delta_0}{T}  &\Delta_0 \ll T 
\\ 1 &T \ll \Delta_0 \end{array} \right.
\end{equation}
in qualitative agreement with the exact result $\langle \sigma_x \rangle =
\tanh ( \Delta_0 / T )$.
\bibliography{References4}

\newcommand{\npb}{Nucl. Phys.}\newcommand{\adv}{Adv.
  Phys.}\newcommand{\epl}{Europhys. Lett.}
\begin{thebibliography}{40}
\expandafter\ifx\csname natexlab\endcsname\relax\def\natexlab#1{#1}\fi
\expandafter\ifx\csname bibnamefont\endcsname\relax
  \def\bibnamefont#1{#1}\fi
\expandafter\ifx\csname bibfnamefont\endcsname\relax
  \def\bibfnamefont#1{#1}\fi
\expandafter\ifx\csname citenamefont\endcsname\relax
  \def\citenamefont#1{#1}\fi
\expandafter\ifx\csname url\endcsname\relax
  \def\url#1{\texttt{#1}}\fi
\expandafter\ifx\csname urlprefix\endcsname\relax\def\urlprefix{URL }\fi
\providecommand{\bibinfo}[2]{#2}
\providecommand{\eprint}[2][]{\url{#2}}

\bibitem[{\citenamefont{Omn\`es}(1992)}]{O92}
\bibinfo{author}{\bibfnamefont{R.}~\bibnamefont{Omn\`es}},
  \bibinfo{journal}{Rev. Mod. Phys.} \textbf{\bibinfo{volume}{64}},
  \bibinfo{pages}{339} (\bibinfo{year}{1992}).

\bibitem[{\citenamefont{Zurek}(2003)}]{Z03}
\bibinfo{author}{\bibfnamefont{W.~H.} \bibnamefont{Zurek}},
  \bibinfo{journal}{Rev. Mod. Phys.} \textbf{\bibinfo{volume}{75}},
  \bibinfo{pages}{715} (\bibinfo{year}{2003}).

\bibitem[{\citenamefont{Caldeira and Leggett}(1981)}]{CL81}
\bibinfo{author}{\bibfnamefont{A.~O.} \bibnamefont{Caldeira}} \bibnamefont{and}
  \bibinfo{author}{\bibfnamefont{A.~J.} \bibnamefont{Leggett}},
  \bibinfo{journal}{Phys. Rev. Lett.} \textbf{\bibinfo{volume}{46}},
  \bibinfo{pages}{211} (\bibinfo{year}{1981}).

\bibitem[{\citenamefont{Caldeira and Leggett}(1983{\natexlab{a}})}]{Cal83}
\bibinfo{author}{\bibfnamefont{A.~O.} \bibnamefont{Caldeira}} \bibnamefont{and}
  \bibinfo{author}{\bibfnamefont{A.~J.} \bibnamefont{Leggett}},
  \bibinfo{journal}{Ann. Phys. (N.Y.)} \textbf{\bibinfo{volume}{149}},
  \bibinfo{pages}{374} (\bibinfo{year}{1983}{\natexlab{a}}).

\bibitem[{\citenamefont{Leggett
  et~al.}(1987{\natexlab{a}})\citenamefont{Leggett, Chakravarty, Dorsey,
  Fisher, Garg, and Zwerger}}]{Letal87}
\bibinfo{author}{\bibfnamefont{A.~J.} \bibnamefont{Leggett}},
  \bibinfo{author}{\bibfnamefont{S.}~\bibnamefont{Chakravarty}},
  \bibinfo{author}{\bibfnamefont{A.~T.} \bibnamefont{Dorsey}},
  \bibinfo{author}{\bibfnamefont{M.~P.~A.} \bibnamefont{Fisher}},
  \bibinfo{author}{\bibfnamefont{A.}~\bibnamefont{Garg}}, \bibnamefont{and}
  \bibinfo{author}{\bibfnamefont{W.}~\bibnamefont{Zwerger}},
  \bibinfo{journal}{Rev. Mod. Phys.} \textbf{\bibinfo{volume}{51}},
  \bibinfo{pages}{1} (\bibinfo{year}{1987}{\natexlab{a}}).

\bibitem[{\citenamefont{Weiss}(1999)}]{W99}
\bibinfo{author}{\bibfnamefont{U.}~\bibnamefont{Weiss}},
  \emph{\bibinfo{title}{Quantum dissipative systems}}
  (\bibinfo{publisher}{World Scientific}, \bibinfo{address}{Singapore},
  \bibinfo{year}{1999}).

\bibitem[{\citenamefont{Osterloh et~al.}(2002)\citenamefont{Osterloh, Amico,
  Falci, and Fazio}}]{Oetal02}
\bibinfo{author}{\bibfnamefont{A.}~\bibnamefont{Osterloh}},
  \bibinfo{author}{\bibfnamefont{L.}~\bibnamefont{Amico}},
  \bibinfo{author}{\bibfnamefont{G.}~\bibnamefont{Falci}}, \bibnamefont{and}
  \bibinfo{author}{\bibfnamefont{R.}~\bibnamefont{Fazio}},
  \bibinfo{journal}{Nature} \textbf{\bibinfo{volume}{416}},
  \bibinfo{pages}{608} (\bibinfo{year}{2002}).

\bibitem[{\citenamefont{Osborne and Nielsen}(2002)}]{ON02}
\bibinfo{author}{\bibfnamefont{T.~J.} \bibnamefont{Osborne}} \bibnamefont{and}
  \bibinfo{author}{\bibfnamefont{M.~A.} \bibnamefont{Nielsen}},
  \bibinfo{journal}{Phys. Rev. A} \textbf{\bibinfo{volume}{66}},
  \bibinfo{pages}{032110} (\bibinfo{year}{2002}).

\bibitem[{\citenamefont{Vidal et~al.}(2004)\citenamefont{Vidal, Palacios, and
  Aslangul}}]{Vidal04}
\bibinfo{author}{\bibfnamefont{J.}~\bibnamefont{Vidal}},
  \bibinfo{author}{\bibfnamefont{G.}~\bibnamefont{Palacios}}, \bibnamefont{and}
  \bibinfo{author}{\bibfnamefont{C.}~\bibnamefont{Aslangul}},
  \bibinfo{journal}{Phys. Rev. A} \textbf{\bibinfo{volume}{70}},
  \bibinfo{pages}{062304} (\bibinfo{year}{2004}).

\bibitem[{\citenamefont{Dusuel and Vidal}(2005)}]{DV05}
\bibinfo{author}{\bibfnamefont{S.}~\bibnamefont{Dusuel}} \bibnamefont{and}
  \bibinfo{author}{\bibfnamefont{J.}~\bibnamefont{Vidal}},
  \bibinfo{journal}{Phys. Rev. B} \textbf{\bibinfo{volume}{71}},
  \bibinfo{pages}{224420} (\bibinfo{year}{2005}).

\bibitem[{\citenamefont{Stauber and Guinea}(2004)}]{SG04}
\bibinfo{author}{\bibfnamefont{T.}~\bibnamefont{Stauber}} \bibnamefont{and}
  \bibinfo{author}{\bibfnamefont{F.}~\bibnamefont{Guinea}},
  \bibinfo{journal}{Phys. Rev. A} \textbf{\bibinfo{volume}{70}},
  \bibinfo{pages}{022313} (\bibinfo{year}{2004}).

\bibitem[{\citenamefont{Anderson et~al.}(1970)\citenamefont{Anderson, Yuval,
  and Hamann}}]{AYH70}
\bibinfo{author}{\bibfnamefont{P.~W.} \bibnamefont{Anderson}},
  \bibinfo{author}{\bibfnamefont{G.}~\bibnamefont{Yuval}}, \bibnamefont{and}
  \bibinfo{author}{\bibfnamefont{D.~R.} \bibnamefont{Hamann}},
  \bibinfo{journal}{Phys. Rev. B} \textbf{\bibinfo{volume}{1}},
  \bibinfo{pages}{4464} (\bibinfo{year}{1970}).

\bibitem[{\citenamefont{Schmid}(1983)}]{S83}
\bibinfo{author}{\bibfnamefont{A.}~\bibnamefont{Schmid}},
  \bibinfo{journal}{Phys. Rev. Lett.} \textbf{\bibinfo{volume}{51}},
  \bibinfo{pages}{1506} (\bibinfo{year}{1983}).

\bibitem[{\citenamefont{Kane and Fisher}(1992)}]{KF92}
\bibinfo{author}{\bibfnamefont{C.~L.} \bibnamefont{Kane}} \bibnamefont{and}
  \bibinfo{author}{\bibfnamefont{M.~P.~A.} \bibnamefont{Fisher}},
  \bibinfo{journal}{Phys. Rev. B} \textbf{\bibinfo{volume}{46}},
  \bibinfo{pages}{15233} (\bibinfo{year}{1992}).

\bibitem[{\citenamefont{Nagaev and B\"uttiker}(2002)}]{NB02}
\bibinfo{author}{\bibfnamefont{K.~E.} \bibnamefont{Nagaev}} \bibnamefont{and}
  \bibinfo{author}{\bibfnamefont{M.}~\bibnamefont{B\"uttiker}},
  \bibinfo{journal}{Europhys. Lett.} \textbf{\bibinfo{volume}{58}},
  \bibinfo{pages}{475} (\bibinfo{year}{2002}).

\bibitem[{\citenamefont{Jordan and B\"uttiker}(2004)}]{JB04}
\bibinfo{author}{\bibfnamefont{A.~N.} \bibnamefont{Jordan}} \bibnamefont{and}
  \bibinfo{author}{\bibfnamefont{M.}~\bibnamefont{B\"uttiker}},
  \bibinfo{journal}{Phys. Rev. Lett.} \textbf{\bibinfo{volume}{92}},
  \bibinfo{pages}{247901} (\bibinfo{year}{2004}).

\bibitem[{\citenamefont{Georges et~al.}(1996)\citenamefont{Georges, Kotliar,
  Krauth, and Rozenberg}}]{GKKR96}
\bibinfo{author}{\bibfnamefont{A.}~\bibnamefont{Georges}},
  \bibinfo{author}{\bibfnamefont{G.}~\bibnamefont{Kotliar}},
  \bibinfo{author}{\bibfnamefont{W.}~\bibnamefont{Krauth}}, \bibnamefont{and}
  \bibinfo{author}{\bibfnamefont{M.}~\bibnamefont{Rozenberg}},
  \bibinfo{journal}{Rev. Mod. Phys.} \textbf{\bibinfo{volume}{68}},
  \bibinfo{pages}{13} (\bibinfo{year}{1996}).

\bibitem[{\citenamefont{Wootters}(1998)}]{W98}
\bibinfo{author}{\bibfnamefont{W.~K.} \bibnamefont{Wootters}},
  \bibinfo{journal}{Phys. Rev. Lett.} \textbf{\bibinfo{volume}{80}},
  \bibinfo{pages}{2245} (\bibinfo{year}{1998}).

\bibitem[{\citenamefont{Verstraete
  et~al.}(2004{\natexlab{a}})\citenamefont{Verstraete, Popp, and
  Cirac}}]{VPC04}
\bibinfo{author}{\bibfnamefont{F.}~\bibnamefont{Verstraete}},
  \bibinfo{author}{\bibfnamefont{M.}~\bibnamefont{Popp}}, \bibnamefont{and}
  \bibinfo{author}{\bibfnamefont{J.~I.} \bibnamefont{Cirac}},
  \bibinfo{journal}{Phys. Rev. Lett.} \textbf{\bibinfo{volume}{92}},
  \bibinfo{pages}{027901} (\bibinfo{year}{2004}{\natexlab{a}}).

\bibitem[{\citenamefont{Vidal et~al.}(2003)\citenamefont{Vidal, Latorre, Rico,
  and Kitaev}}]{Vetal03}
\bibinfo{author}{\bibfnamefont{G.}~\bibnamefont{Vidal}},
  \bibinfo{author}{\bibfnamefont{J.~I.} \bibnamefont{Latorre}},
  \bibinfo{author}{\bibfnamefont{E.}~\bibnamefont{Rico}}, \bibnamefont{and}
  \bibinfo{author}{\bibfnamefont{A.}~\bibnamefont{Kitaev}},
  \bibinfo{journal}{Phys. Rev. Lett.} \textbf{\bibinfo{volume}{90}},
  \bibinfo{pages}{227902} (\bibinfo{year}{2003}).

\bibitem[{\citenamefont{Verstraete
  et~al.}(2004{\natexlab{b}})\citenamefont{Verstraete, Martin-Delgado, and
  Cirac}}]{VMC04}
\bibinfo{author}{\bibfnamefont{F.}~\bibnamefont{Verstraete}},
  \bibinfo{author}{\bibfnamefont{M.~A.} \bibnamefont{Martin-Delgado}},
  \bibnamefont{and} \bibinfo{author}{\bibfnamefont{J.~I.} \bibnamefont{Cirac}},
  \bibinfo{journal}{Phys. Rev. Lett.} \textbf{\bibinfo{volume}{92}},
  \bibinfo{pages}{087201} (\bibinfo{year}{2004}{\natexlab{b}}).

\bibitem[{\citenamefont{Eisert and Plenio}(2002)}]{EP02}
\bibinfo{author}{\bibfnamefont{J.}~\bibnamefont{Eisert}} \bibnamefont{and}
  \bibinfo{author}{\bibfnamefont{M.~B.} \bibnamefont{Plenio}},
  \bibinfo{journal}{Phys. Rev. Lett.} \textbf{\bibinfo{volume}{89}},
  \bibinfo{pages}{137902} (\bibinfo{year}{2002}).

\bibitem[{\citenamefont{Audenaert et~al.}(2002)\citenamefont{Audenaert, Eisert,
  Plenio, and Werner}}]{AEPW02}
\bibinfo{author}{\bibfnamefont{K.}~\bibnamefont{Audenaert}},
  \bibinfo{author}{\bibfnamefont{J.}~\bibnamefont{Eisert}},
  \bibinfo{author}{\bibfnamefont{M.~B.} \bibnamefont{Plenio}},
  \bibnamefont{and} \bibinfo{author}{\bibfnamefont{R.~F.}
  \bibnamefont{Werner}}, \bibinfo{journal}{Phys. Rev. A}
  \textbf{\bibinfo{volume}{66}}, \bibinfo{pages}{042327}
  (\bibinfo{year}{2002}).

\bibitem[{\citenamefont{Paladino et~al.}(2002)\citenamefont{Paladino, Faoro,
  Falci, and Fazio}}]{PFFF02}
\bibinfo{author}{\bibfnamefont{E.}~\bibnamefont{Paladino}},
  \bibinfo{author}{\bibfnamefont{L.}~\bibnamefont{Faoro}},
  \bibinfo{author}{\bibfnamefont{G.}~\bibnamefont{Falci}}, \bibnamefont{and}
  \bibinfo{author}{\bibfnamefont{R.}~\bibnamefont{Fazio}},
  \bibinfo{journal}{Phys. Rev. Lett.} \textbf{\bibinfo{volume}{88}},
  \bibinfo{pages}{228304} (\bibinfo{year}{2002}).

\bibitem[{\citenamefont{Falci et~al.}(2004)\citenamefont{Falci, D'Arrigo,
  Mastellone, and Paladino}}]{FDMP04}
\bibinfo{author}{\bibfnamefont{G.}~\bibnamefont{Falci}},
  \bibinfo{author}{\bibfnamefont{A.}~\bibnamefont{D'Arrigo}},
  \bibinfo{author}{\bibfnamefont{A.}~\bibnamefont{Mastellone}},
  \bibnamefont{and} \bibinfo{author}{\bibfnamefont{E.}~\bibnamefont{Paladino}},
  \bibinfo{journal}{Phys. Rev. A} \textbf{\bibinfo{volume}{70}},
  \bibinfo{pages}{040101(R)} (\bibinfo{year}{2004}).

\bibitem[{\citenamefont{Caldeira and Leggett}(1983{\natexlab{b}})}]{Cal83b}
\bibinfo{author}{\bibfnamefont{A.~O.} \bibnamefont{Caldeira}} \bibnamefont{and}
  \bibinfo{author}{\bibfnamefont{A.~J.} \bibnamefont{Leggett}},
  \bibinfo{journal}{Physica A} \textbf{\bibinfo{volume}{121}},
  \bibinfo{pages}{587} (\bibinfo{year}{1983}{\natexlab{b}}).

\bibitem[{\citenamefont{Hakim and Ambegaokar}(1985)}]{Hak85}
\bibinfo{author}{\bibfnamefont{V.}~\bibnamefont{Hakim}} \bibnamefont{and}
  \bibinfo{author}{\bibfnamefont{V.}~\bibnamefont{Ambegaokar}},
  \bibinfo{journal}{Phys. Rev. A} \textbf{\bibinfo{volume}{32}},
  \bibinfo{pages}{423} (\bibinfo{year}{1985}).

\bibitem[{\citenamefont{Bray and Moore}(1982)}]{Bra82}
\bibinfo{author}{\bibfnamefont{A.~J.} \bibnamefont{Bray}} \bibnamefont{and}
  \bibinfo{author}{\bibfnamefont{M.~A.} \bibnamefont{Moore}},
  \bibinfo{journal}{Phys. Rev. Lett.} \textbf{\bibinfo{volume}{49}},
  \bibinfo{pages}{1545} (\bibinfo{year}{1982}).

\bibitem[{\citenamefont{Chakravarty}(1982)}]{Cha82}
\bibinfo{author}{\bibfnamefont{S.}~\bibnamefont{Chakravarty}},
  \bibinfo{journal}{Phys. Rev. Lett.} \textbf{\bibinfo{volume}{49}},
  \bibinfo{pages}{681} (\bibinfo{year}{1982}).

\bibitem[{\citenamefont{Kosterlitz}(1976)}]{K76}
\bibinfo{author}{\bibfnamefont{J.~M.} \bibnamefont{Kosterlitz}},
  \bibinfo{journal}{Phys. Rev. Lett.} \textbf{\bibinfo{volume}{37}},
  \bibinfo{pages}{1577} (\bibinfo{year}{1976}).

\bibitem[{\citenamefont{Spohn and D\"umcke}(1985)}]{Spohn85}
\bibinfo{author}{\bibfnamefont{H.}~\bibnamefont{Spohn}} \bibnamefont{and}
  \bibinfo{author}{\bibfnamefont{R.}~\bibnamefont{D\"umcke}},
  \bibinfo{journal}{J. Stat. Phys.} \textbf{\bibinfo{volume}{41}},
  \bibinfo{pages}{389} (\bibinfo{year}{1985}).

\bibitem[{\citenamefont{Kehrein and Mielke}(1996)}]{KM96}
\bibinfo{author}{\bibfnamefont{S.~K.} \bibnamefont{Kehrein}} \bibnamefont{and}
  \bibinfo{author}{\bibfnamefont{A.}~\bibnamefont{Mielke}},
  \bibinfo{journal}{Phys. Lett. A} \textbf{\bibinfo{volume}{219}},
  \bibinfo{pages}{313} (\bibinfo{year}{1996}).

\bibitem[{\citenamefont{Bulla et~al.}(2003)\citenamefont{Bulla, Tong, and
  Vojta}}]{Bulla03}
\bibinfo{author}{\bibfnamefont{R.}~\bibnamefont{Bulla}},
  \bibinfo{author}{\bibfnamefont{N.~H.} \bibnamefont{Tong}}, \bibnamefont{and}
  \bibinfo{author}{\bibfnamefont{M.}~\bibnamefont{Vojta}},
  \bibinfo{journal}{Phys. Rev. Lett.} \textbf{\bibinfo{volume}{91}},
  \bibinfo{pages}{170601} (\bibinfo{year}{2003}).

\bibitem[{\citenamefont{Stauber and Mielke}(2002)}]{SM02}
\bibinfo{author}{\bibfnamefont{T.}~\bibnamefont{Stauber}} \bibnamefont{and}
  \bibinfo{author}{\bibfnamefont{A.}~\bibnamefont{Mielke}},
  \bibinfo{journal}{Phys. Lett. A} \textbf{\bibinfo{volume}{305}},
  \bibinfo{pages}{275} (\bibinfo{year}{2002}).

\bibitem[{\citenamefont{Guinea}(1985)}]{G85}
\bibinfo{author}{\bibfnamefont{F.}~\bibnamefont{Guinea}},
  \bibinfo{journal}{Phys. Rev. B} \textbf{\bibinfo{volume}{32}},
  \bibinfo{pages}{4486} (\bibinfo{year}{1985}).

\bibitem[{\citenamefont{Costi and Kieffer}(1996)}]{Costi96}
\bibinfo{author}{\bibfnamefont{T.~A.} \bibnamefont{Costi}} \bibnamefont{and}
  \bibinfo{author}{\bibfnamefont{C.}~\bibnamefont{Kieffer}},
  \bibinfo{journal}{Phys. Rev. Lett.} \textbf{\bibinfo{volume}{76}},
  \bibinfo{pages}{1683} (\bibinfo{year}{1996}).

\bibitem[{\citenamefont{Vojta et~al.}(2005)\citenamefont{Vojta, Tong, and
  Bulla}}]{VTB05}
\bibinfo{author}{\bibfnamefont{M.}~\bibnamefont{Vojta}},
  \bibinfo{author}{\bibfnamefont{N.~H.} \bibnamefont{Tong}}, \bibnamefont{and}
  \bibinfo{author}{\bibfnamefont{R.}~\bibnamefont{Bulla}},
  \bibinfo{journal}{Phys. Rev. Lett.} \textbf{\bibinfo{volume}{94}},
  \bibinfo{pages}{070604} (\bibinfo{year}{2005}).

\bibitem[{\citenamefont{Leggett
  et~al.}(1987{\natexlab{b}})\citenamefont{Leggett, Chakravarty, Dorsey,
  Fisher, Garg, and Zwerger}}]{Leggett87}
\bibinfo{author}{\bibfnamefont{A.~J.} \bibnamefont{Leggett}},
  \bibinfo{author}{\bibfnamefont{S.}~\bibnamefont{Chakravarty}},
  \bibinfo{author}{\bibfnamefont{A.~T.} \bibnamefont{Dorsey}},
  \bibinfo{author}{\bibfnamefont{M.~P.~A.} \bibnamefont{Fisher}},
  \bibinfo{author}{\bibfnamefont{A.}~\bibnamefont{Garg}}, \bibnamefont{and}
  \bibinfo{author}{\bibfnamefont{W.}~\bibnamefont{Zwerger}},
  \bibinfo{journal}{Rev. Mod. Phys} \textbf{\bibinfo{volume}{59}},
  \bibinfo{pages}{1} (\bibinfo{year}{1987}{\natexlab{b}}).

\bibitem[{\citenamefont{Anderson and Yuval}(1971)}]{AY71}
\bibinfo{author}{\bibfnamefont{P.~W.} \bibnamefont{Anderson}} \bibnamefont{and}
  \bibinfo{author}{\bibfnamefont{G.}~\bibnamefont{Yuval}}, \bibinfo{journal}{J.
  Phys. C: Cond. Mat.} \textbf{\bibinfo{volume}{4}}, \bibinfo{pages}{607}
  (\bibinfo{year}{1971}).

\bibitem[{\citenamefont{Cardy}(1981)}]{C81}
\bibinfo{author}{\bibfnamefont{J.~L.} \bibnamefont{Cardy}},
  \bibinfo{journal}{J. Phys. A: Math. and Gen.} \textbf{\bibinfo{volume}{14}},
  \bibinfo{pages}{1407} (\bibinfo{year}{1981}).

\end{thebibliography}
\end{document}